\documentstyle[preprint,psfig,aps]{revtex}
\tightenlines
\begin{document}
\draft
\title{Quantum Hall Effect induced by electron-electron interaction
in disordered GaAs layers with 3D spectrum}
\author{S.S. Murzin$^{\text{1,2}}$, I. Claus$^{\text{2}}$, A.G.M. Jansen$^{\text{2}}$,
N.T. Moshegov$^{\text{3}}$, A.I. Toropov$^{\text{3}}$, and K. Eberl$^{\text{4}}$}
\address{$^{\text{1}}$Institute of Solid State Physics RAS, 142432, Chernogolovka,
Moscow distr., Russia\\
$^{\text{2}}$Grenoble High Magnetic Field Laboratory, Max-Plank-Institut
f\"ur Festk\"orperforschung and Centre Nationalde la Recherche Scientifique,
B.P.166, F-38042, Grenoble Cedex 9, France \\
$^{\text{3}}$Institute of Semiconductor Physics RAS, 630090, Novosibirsk,
Russia\\
$^{\text{4}}$Max-Plank-Institut f\"ur Festk\"orperforschung, Postfach 800
665 D-70569, Stuttgart, Germany}
\date{\today}
\maketitle

\begin{abstract}
It is shown that the observed Quantum Hall Effect in epitaxial layers of
heavily doped n-type GaAs with thickness ($50-140$~nm) larger the mean
free path of the conduction electrons ($15-30$~nm) and, therefore, with a
three-dimensional single-particle spectrum is induced by the
electron-electron interaction. The Hall resistance $R_{xy}$ of the thinnest
sample reveals a  wide plateau at small activation
energy $E_a=0.4$~K found in the temperature dependence of the transverse
resistance $R_{xx}$.  The different minima in the transverse conductance
$G_{xx}$ of the different samples show a universal
temperature dependence
(logarithmic in a large
range of rescaled temperatures $T/T_0$) which is reminiscent of
electron-electron-interaction effects in coherent diffusive transport.
\end{abstract}


\narrowtext

For a two-dimensional (2D) electron system it is well known that the
discrete Landau level spectrum of an electron in a quantizing magnetic
field leads to quantized magnetotransport \cite{Prange}.  In the Integer
Quantum Hall Effect (IQHE) quantized plateaus are observed in the Hall
conductance and Hall resistance ($G_{xy}=1/R_{xy}=ie^2/h$ with integer
$i$) together with zero-valued transverse condunctance $G_{xx}$ and
transverse resistance $R_{xx}$ at magnetic fields where the Fermi level
lies at localized states in between the Landau levels.  In this work,
experimental evidence is given that an IQHE can also result from
electron-electron interaction effects in a disordered layer with a
continuous three-dimensional (3D) single-particle spectrum of the charge
carriers.

Recently, a quantization of the Hall conductance has been observed in
disordered epitaxial layers of heavily doped n-type GaAs  with a thickness
around 100~nm \cite{mjl}. At 4.2~K, the magnetoresistance of these samples
shows the typical behavior of bulk material with weak Shubnikov-de Haas
oscillations for increasing field $B$ and a strong monotonous upturn in
the extreme quantum limit (EQL) where only the lowest Landau level is
occupied.  Below 1~K, the magnetic-field dependence of the Hall
conductance $G_{xy}$ shows steps at quantized values $ie^2/h$ with i=2, 4,
and 6 together with pronounced minima in the transverse conductance
$G_{xx}$ in the extreme quantum limit of the applied magnetic field, while
the amplitude of the usual Shubnikov-de Haas oscillations below the EQL do
not depend on temperature $T$ below 4~K.

Ignoring quantum electron-electron correlation effects, the electronic
system in the investigated GaAs layers has a 3D Landau-level spectrum
without variations of the density of states due to dimensional
quantization across the layer.  The 3D character has been confirmed in
experiments in a tilted magnetic field \cite{mcj}.  The magnetoresistance
$R_{xx}$ at $T=4.2$~K shows only an insignificant change by rotating the
field away from the vertical orientation.  At low temperatures, the minima
in $R_{xx}$ and $dR_{xy}/dB$ shift much slower to higher fields and are
suppressed much faster as compared to the expected angular dependence for
a two-dimensional system \cite{mcj}.  We note however, that below 3~-~4~K
the characteristic diffusion lengths,
$L_{\varphi}=(D_{zz}\tau_{\varphi})^{1/2}$ and $L_{T}=(D_{zz}\hbar
/k_BT)^{1/2}$, for coherent transport increase to values larger than $d$,
and the system becomes 2D for coherent phenomena in the diffusive
transport ($D_{zz}$ is the diffusion coefficient of  electrons along the
magnetic field, $\tau_{\phi}$ is the phase relaxation time).

On different Si-doped GaAs layers (thicknesses 50~-~140~nm and
charge-carrier concentrations 0.8~-~$1.5 \times 10^{17}$~cm$^{-3}$) the
temperature dependence of the QHE has been investigated.  Both the
observed excitation gap and universal scaling behavior in the temperature
dependence of the conductance reveal the importance of electron-electron
interaction effects for the explanation of the phenomenon.

The samples (see Table 1) used were prepared by molecular-beam epitaxy.
Samples 1-3 are  described in reference \cite{mjl}. Sample 4 has the same
structure as samples 2 and 3 but was grown at a somewhat  higher
temperature than the optimal one.  Samples 5-7 have a slightly different
structure: on a GaAs (100) substrate were successively grown an undoped
GaAs layer (0.1~$\mu $m), a periodic structure 30~$\times $
GaAs/AlGaAs(10/10 nm), an undoped GaAs layer (0.5$\mu $m), the heavily
Si-doped GaAs of nominal thickness $d=50, 100$ and $140$~nm and donor(Si)
concentrations $1.5\times 10^{17}$ cm$^{-3}$ and a cap layer (0.5 $\mu
$m). Hall bar geometries of width 0.2~mm and length 2.8~mm were etched out
of the wafers. A phase sensitive ac-technique was used for the
magnetotransport measurements down to 50~mK.  In the experiments the
applied magnetic field up to 23~T was directed perpendicular to the
layers.

The bulk density of electrons $n$ in a heavily doped layer has been
determined from the periodicity of Shubnikov-de Haas oscillations in the
transverse resistance $R_{xx}$ with the results as listed in Table~1 for
all investigated samples. The mobilities $\mu$ have been determined from
the zero-magnetic-field resistance and the Hall resistance $R_{xy}$ in the
linear region of weak magnetic fields at $T=4.2$~K (see Table~1). They are
close to the mobilities found for bulk material with comparable electron
densities\cite{abm} except the mobility of sample 4 with a somewhat lower
value. The average thickness $d_{ex}$ of the conducting layers (see
Table~1) has been obtained by dividing the electron density $N_s$ per area
unit, as determined from the Hall measurements, by the bulk density $n$.
The values of $d_{ex}$ for the investigated samples are somewhat smaller
than the nominal thicknesses due to the decreasing  electron density near
the interfaces.

\begin{figure}[t]
\psfig{file=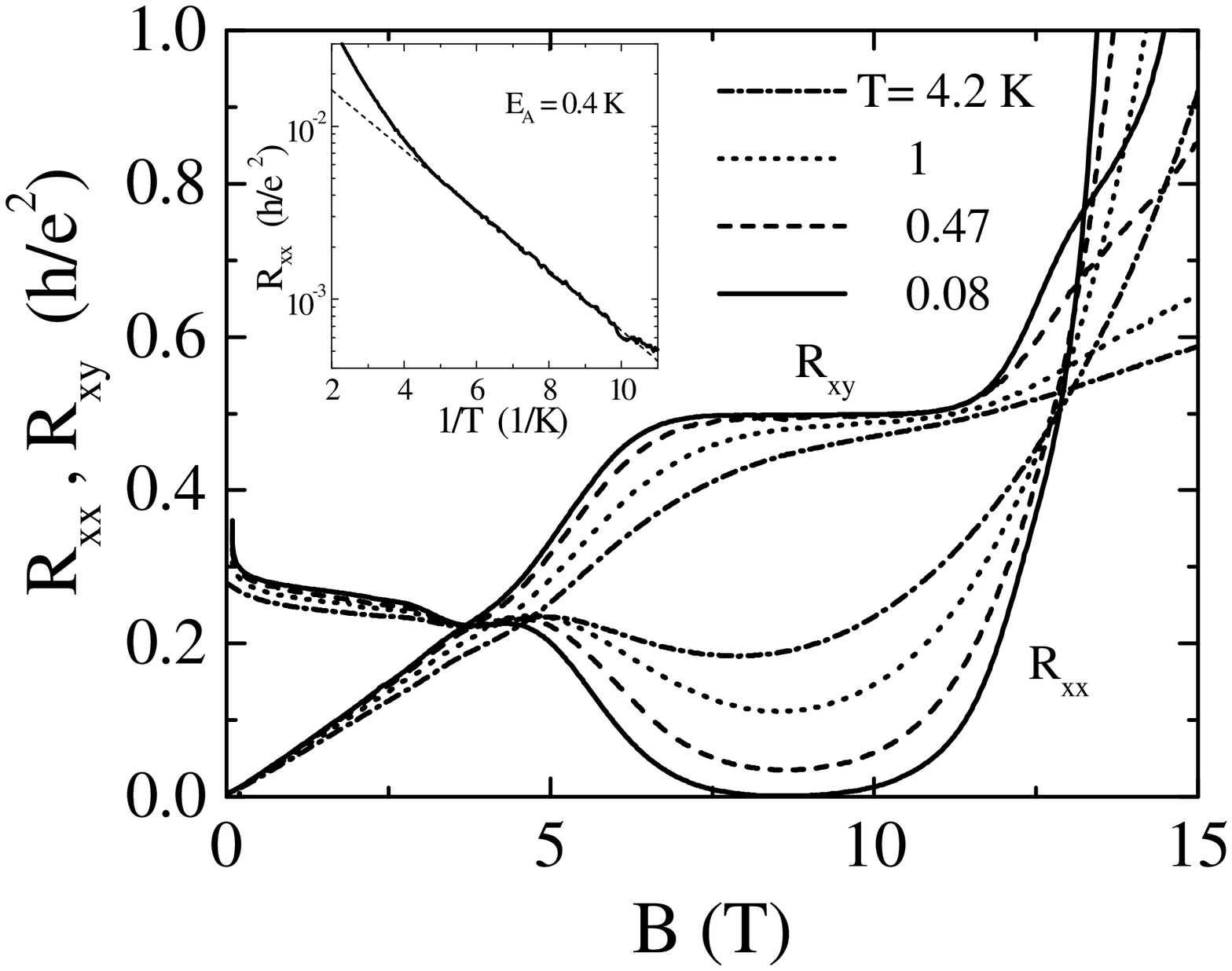,width=16cm,clip=}

~

\caption{Magnetic field dependence of the Hall resistance $R_{xy}$ and
transverse resistance $R_{xx}$ (per square) for sample~5 in a magnetic
field perpendicular to the GaAs layer at two temperatures. $R_{xy}$ reveals
a wide plateau at $h/2e^2$ from  $B=7$~T up to 11 T at low temperatures with
a deep minimum in $R_{xx}$. The insert shows an activated behavior $R_{xx}
\propto \exp (-E_A/T)$ with $E_A = 0.4$~K for temperature dependence in
minimum at $B=8.65$~T.}
\label{resist}
\end{figure}

In Fig.\ref{resist} the magnetotransport data of the Hall resistance
$R_{xy}$ and transverse $R_{xx}$ (per square) have been plotted for the
thinnest sample~5 at temperatures below 4.2~K.
The Hall resistance $R_{xy}$
reveals a wide plateau from $B=7$~T up to 11~T
at the lowest temperatures with
the value $R_{xy}=h/2e^2$ (i.e. $i=2$), while $R_{xx}$ reveals a deep
minimum.  For samples~1,2, 4 and 6 the resistance minima at the Hall
plateaus do not descend to zero, neither are the Hall plateaus completely
flat (see the data in ref.\cite{mjl}).  Such a behavior is also seen in
the QHE for 2D systems in case of a not complete opening of the mobility
gap in between the Landau levels.  For samples~3 and 7, only weak
temperature-dependent variations are observed at the Hall plateaus with
$i=4, 6, 8$ and  $i=2, 4, 6, 8$, respectively.

\begin{figure}[t]
\psfig{file=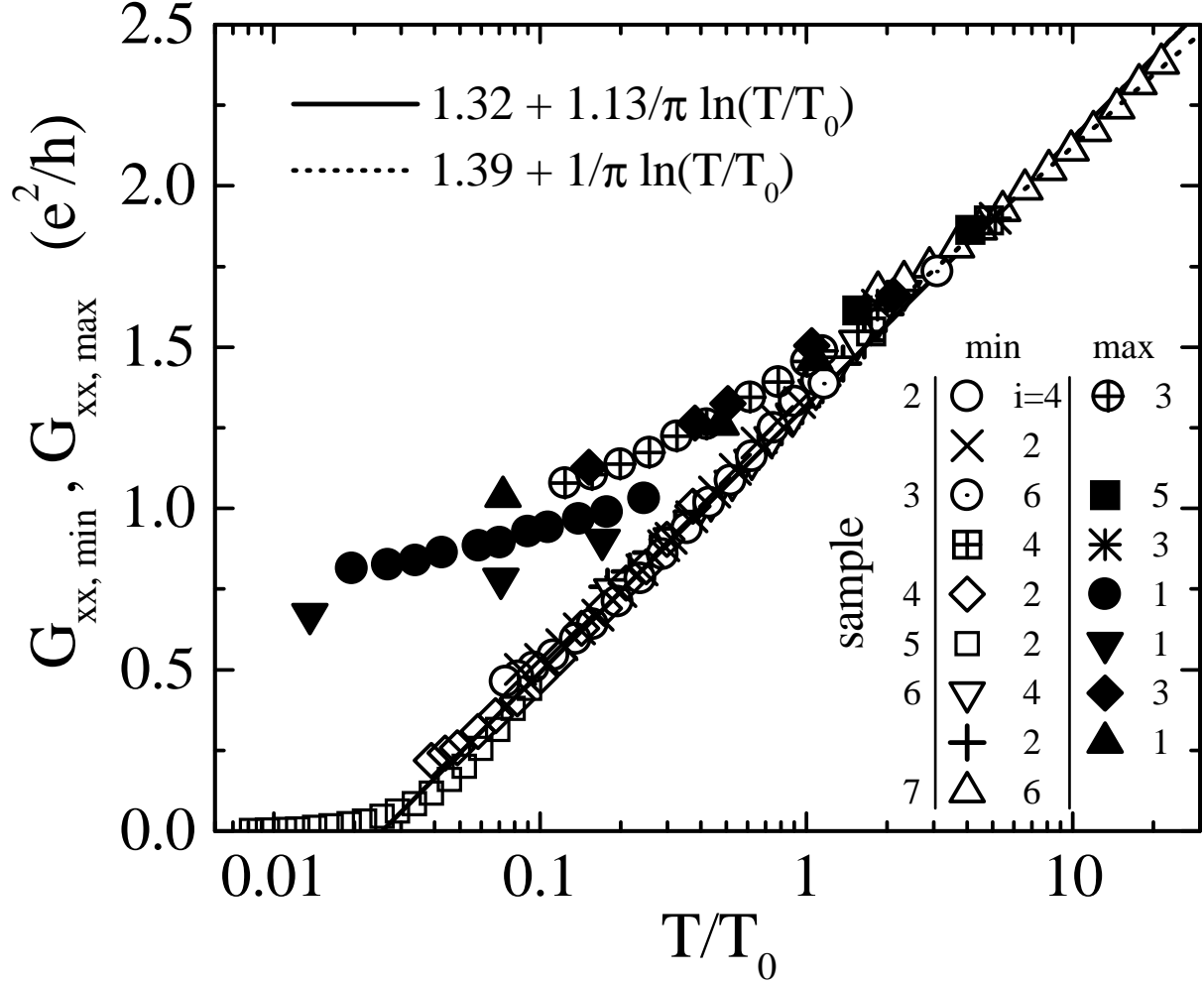,width=16cm,clip=}

~

\caption{Conductance $G_{xx,min}$ at the minima and $G_{xx,max}$ at the
maxima as a function of the normalized temperature $T/T_0$ for different
minima and maxima in different samples.
The straight solid and dash lines correspond to the dependences
$G_{xx,min}=e^2/h[1.32+1.13/\pi~ \ln{(T/T_0)}]$
$G_{xx,min}=e^2/h[1.39+1/\pi~ \ln{(T/T_0)}]$, respectively,
at the minima. In the legend, the first  column indicates the sample number, the
second gives the data symbols for the minima at even quantum numbers $i$
($G_{xy}=ie^2/h$) and the third the data symbols for the maxima denoted by
odd numbers. The values of the magnetic field for the minima are, successively,
11.8, 17.7; 14, 20; 7.1; 8.65; 11.6, 16.9; 10.4~T and for the maxima 14.3; 16.8,
22.5; 9.3; 12.6, 14, 19~T .}
\label{scaling}
\end{figure}

\begin{figure}[t]
\psfig{file=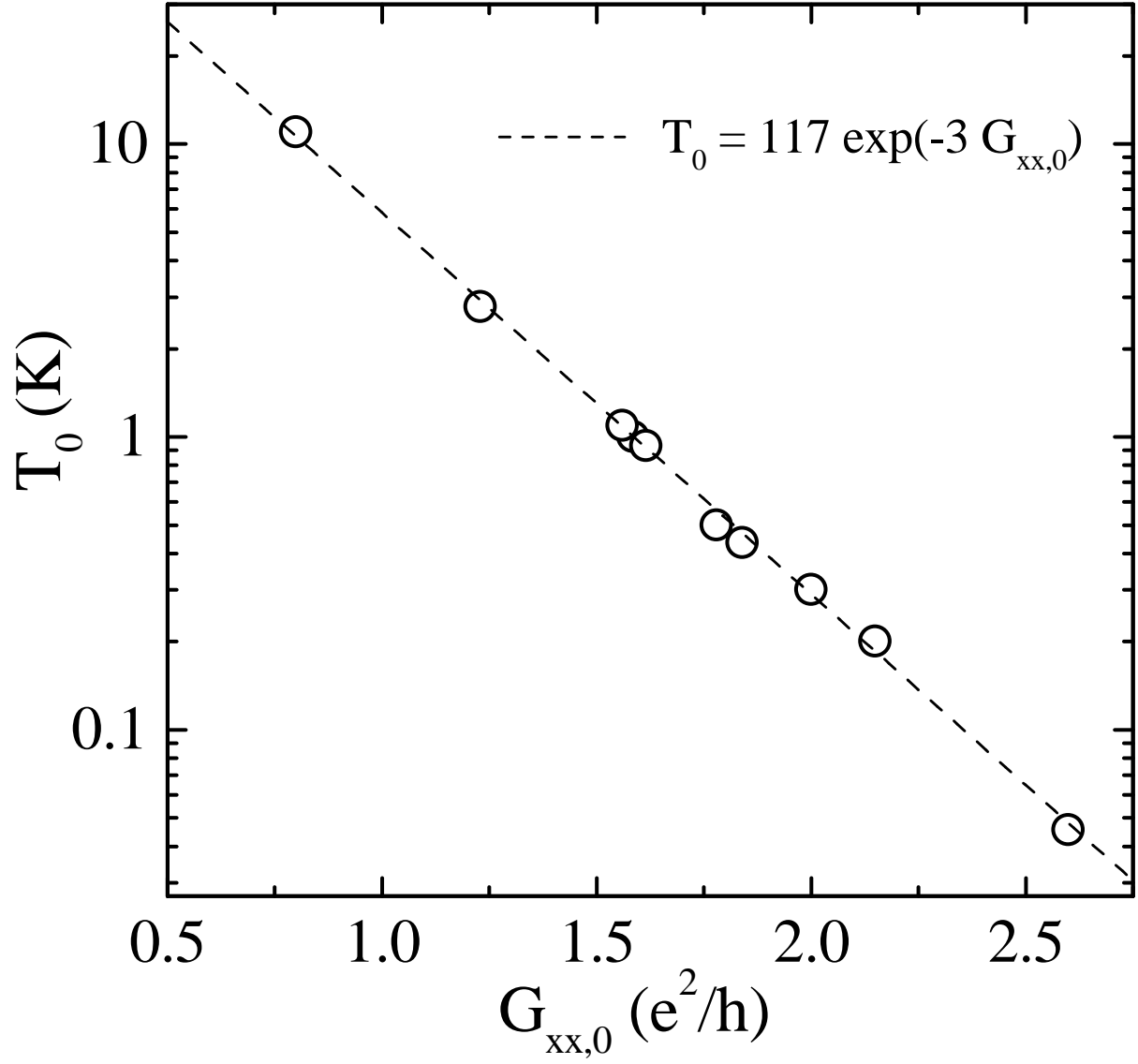,width=16cm,clip=}

~

\caption{Logarithm of the scaling factor $T_0$ as a function of the
conductance $G_{xx,0}$ at $T=4.2$~K as deduced from the data presentation
in Fig.(\ref{scaling}).}
\label{T0}
\end{figure}

In Fig.\ref{scaling} we have summarized the temperature dependence of the
conductance for all samples investigated (except sample 1
to avoid too densily packed data points)
by plotting the conductance $G_{xx,min}$ (per
square) at the minima  as a function of $T/T_0$ for temperatures from
0.07-0.1~K up to 1~K at the different Hall-conductance plateaus with
$i=2$, 4, or 6. For the minimum corresponding to  $i=2$ of sample 2, the
value $T_0$ was taken to be 1~K.  For the other temperature dependencies,
the scaling parameter $T_0$ was chosen such in order to bring all the
$G_{xx,min}(T)$ curves together.  Over a large range of $T/T_0$ the
temperature dependence of the conductance $G_{xx,min}$ is close to
logarithmic and described by
\begin{equation}
G_{xx,min}(T)=e^2/h \left(1.32+\frac{1.13}{\pi} \ln{\frac{T}{T_0}}\right) .
\label{Gmin}
\end{equation}
Above 1~K, the temperature dependences deviate from the logarithm one with
reduced values near 4~K.  The scaling factors $T_0$ depend on  the
conductance $G_{xx,0}$ at $T=4.2$~K in the applied magnetic field, at
which temperature interference effects are expected to be unimportant. As
shown in Fig.\ref{T0}, the values of $T_0$ are well described by the
dependence
\begin{equation}
T_0=117 \exp (-3 G_{xx,0} h/e^2)~[{\rm K}]~.
\label{T_0}
\end{equation}
Note, that Eq.(\ref{T_0}) results automatically from the procedure of
matching logarithmic curves like Eq.(\ref{Gmin}).  If the logarithmic
dependences of Eq.(\ref{Gmin}) would continue up to 4.2~K, the numeric
coefficients in Eq.(\ref{T_0}) would be $4.2 \exp (1.32 \pi /1.13)=165$
and $\pi/1.13= 2.78$ instead of 117 and 3, respectively.

The obtained logarithmic temperature dependence of the conductance is well
known from the small corrections to the conductance as a result of
electron-electron interaction effects in the case of coherent diffusive
transport in 2D~\cite{AA}.  In our case the temperature dependences in
the minima are close
to logarithmic even for strong changes of the conductance.
The coefficient in the logarithmic term of Eq.(\ref{Gmin})
is somewhat larger
than the maximum value $1/\pi$ in the perturbation theory of small
corrections \cite{AA}.  Only for the thickest sample~7 the changes in the
minima of $G_{xx}$ are small (about 5$\%$) with a better description of
the logarithmic temperature dependence using the coefficient $1/\pi$.
The temperature dependence of the conductances $G_{xx}$ and
$G_{xy}$ for sample~2 and 3, in lower magnetic fields ($B<8$~T) where
$G_{xx,0}>4 e^2/h$,
show a good description in terms of quantum corrections
of electron-electron-interaction effects \cite{m98}.

There is no theory of electron-electron interactions in the diffusive
transport beyond the limit of small corrections.  However, we note that a
logarithmic temperature dependence with a strong decrease of $G_{xx}$ was
observed in pure 2D GaAs/AlGaAs heterostructures in the maxima of
$G_{xx}$~\cite{p}.  It was interpreted as a result of electron-electron
interaction effects~\cite{p,g}.

Sample~5, showing the strongest Quantum Hall Effect structures, has an
activated behavior of the resistance $R_{xx} \propto \exp{(-E_A/T)}$ at
temperatures below 0.2~K ($T/T_0 < 0.02$ in Fig.2) with an activation
energy $E_A=0.4$~K (see insert Fig.\ref{resist}).
The large width of the Hall
plateau (from $B =7$~T up to 11~T) in sample 5
compared to the small gap in the activated behavior of the resistance
minimum indicates that there is a Coulomb gap at low temperatures which
moves together with the Fermi level when the magnetic field changes.  All
energy scales in the single-particle approach, except spin-splitting, are
much larger than $E_A = 0.4$~K.  The Fermi energy $E_F$ in zero magnetic
equals 140~K, the scattering-induced  broadening $\hbar/\tau=100$~K, the
cyclotron energy $\hbar \omega_c=180$~K, and the distance $3 (\pi
\hbar/d)^2/2 m$ between the lowest levels of dimensional quantization in a
pure layer of the same 50-nm thickness equals 80~K. The spin-splitting can
hardly be important for even $i$. The narrow gap can be induced only by
electron-electron correlations with a large correlation length. In the
case of a narrow single-particle mobility gap non-attached to the Fermi
level, the Hall plateau would be expected to be very narrow because $E_F$
changes about 2 times in the field range 7-11~T.

In Fig.\ref{scaling} we have also plotted the conductance $G_{xx,max}$ for
the maxima in between the quantized minima
(denoted by odd numbers in the legenda)
and between the minima with $i=2$
and an insulator state at higher fields (denoted by 1).
Using the same Eq.(\ref{T_0}) for the calculation of the rescaled
$T/T_0$ as for the minima,
the data for the maxima also show a
universal behavior for the samples with $G_{xx,0} > 1.5 e^2/h$.  The
data for the maxima have somewhat lower values for the samples with
$G_{xx,0} < 1.3 e^2/h$ (samples 4 and 5) .

The possibility of a  Hall-conductance quantization in a system with a
continuous spectrum  has been suggested by Khmel'nitski\u{\i}~\cite{Khm}
in the frame of non-interacting quasi particles. According to this
approach, in a layer (film) with high-temperature conductances in a
magnetic field such that $G_{xx,0}$ and $G_{xy,0} \gg e^2/h$, the
conductance $G_{xx}$ decreases due to quantum coherent (localization)
effects when the temperature goes down. At the beginning the decrease is
described by second order quantum corrections~\cite{Huck,m98}. When
$G_{xx}$ becomes of order $e^2/h$, the quantization of $G_{xy}$
should develop.

However, in a  layer with $G_{xx,0}$ and $G_{xy,0} \gg e^2/h$ the
quantum corrections to $G_{xx}$ due to electron-electron correlations (of
the first order) are much larger than that due to the single-particle localization
effects (of the second order). A calculation of these latter single-particle
corrections to the conductivity of our samples~\cite{m98} results in a much
weaker temperature dependence of $G_{xx}$ than experimentally observed even
in the regions of magnetic fields where the QHE structure is observed. Since
$\log T$ diverges at $T \to 0$, one may expect that the enhancement of interaction
effects with decreasing temperature results in a Coulomb gap at the Fermi level
with a vanishing dissipative conductance $G_{xx}$ in some ranges of the magnetic
field. In this case the Hall conductance should be quantized according to
Laughlin's argument~\cite{Laugh},  which has also validity for a three dimensional
system~\cite{Ao}. Since there is no designated value of $i$,   $G_{xy}$ will
apparently approach the most nearby quantized value as temperature decreases.
In between, transitional regions should exist where $G_{xy}$ is not quantized and,
correspondingly,  $G_{xx}$ approaches a constant value. Our experimental results
indicate that this scenario is realized.

In summary,  the large width of the Hall plateau at a small excitation gap
is a strong evidence that the novel phenomenon of an IQHE in disordered
GaAs layers with a three-dimensional single-particle spectrum  is induced
by electron-electron interactions.
The observed logarithmic temperature dependence in
the universal $T/T_0$ dependence of the conductance minimum $G_{xx,min}$,
with $T_0$ only depending on the high-temperature conductance
$G_{xx,0}$,
is a continuation of the small logarithmic quantum corrections
of electron-electron interactions in the diffusive transport.

This work is supported by RFBR-PICS (grant 98-02-22037) and
the program "Physics of Solid State Nanostructures" (grant 1-085/4).
We would like to thank B. Hammer for her help in the
preparation of the samples.

\begin{table}
\begin{tabular}{cccccccc}
Sample & $n$ (cm$^{-3}$) & $\mu$ (cm$^2$/Vs) & $d_n$ (nm) & $d_{ex}$ (nm)
&i&$G_{xx,0}(e^2/h)$ \\
\tableline
1 &$ 0.8 \times 10^{17}$&1900 & 100 & 70 & 2 & 1.56 \\
2 &$1.8 \times 10^{17}$ & 2400 & 100 & 70 & 2 &1.62 & \\
 & & & & & 4 &1.59 \\
3 & $2.5 \times 10^{17}$ & 2500 & 100 & 80 & 4 & 2.15 \\
  & & & & & 6 & 2 \\
4 & $0.85 \times 10^{17}$ & 1000 & -  & 50 & 2 &1.23 \\
5 & $1 \times 10^{17}$ & 1900 & 50 & 50 & 2 & 0.8 \\
6 & $1.6 \times 10^{17}$ & 2600 & 100 & 80 & 4 & 1.84 \\
 & & & & &6 &1.78 & \\
7 & $1.6 \times 10^{17}$ & 2600 & 140 & 110 & 6 &2.6
\end{tabular}
\end{table}

\end{document}